\documentclass[12pt,notitlepage]{article}
\usepackage{amsmath,amssymb}
\usepackage{graphicx}

\usepackage{color}

\def\ignore#1{{}}

\newcounter{sxn}

\newcounter{axn}

\newdimen\mybaselineskip
\newcommand{\beeq}{\begin{equation}}
\newcommand{\eneq}{\end{equation}}
\newcommand{\beqn}{\begin{eqnarray}}
\newcommand{\eeqn}{\end{eqnarray}}

\newcommand{\ba}{\begin{array}}
\newcommand{\ea}{\end{array}}

\newcommand{\be}{\begin{equation}}
\newcommand{\ee}{\end{equation}}
\newcommand{\bea}{\begin{eqnarray}}
\newcommand{\eea}{\end{eqnarray}}
\newcommand{\beal}{\setcounter{letter}{1} \begin{eqnarray}}
\newcommand{\eeal}{\addtocounter{equation}{1} \end{eqnarray}}

\newcommand{\larrow}{\,\,\,\,\hbox to 30pt{\rightarrowfill}
\,\,\,\,}
\newcommand{\slarrow}{\,\,\,\hbox to 20pt{\rightarrowfill}
\,\,\,}

\def\la{\raise.16ex\hbox{$\langle$}\lower.16ex\hbox{}  }
\def\ra{\, \raise.16ex\hbox{$\rangle$}\lower.16ex\hbox{} }

\def\psibar{ \psi \kern-.65em\raise.6em\hbox{$-$} \lower.6em\hbox{} }
\def\psibarb{ \psi \kern-.65em\raise.6em\hbox{$-$}  }

\begin{document}

\thispagestyle{empty}





\begin{center}  
{\LARGE \bf   Scalar Perturbations of a Single-Horizon Regular Black Hole}

\vspace{1cm}

{\bf  Ramin G.~Daghigh$^{1}$, Michael D.\ Green$^2$, Jodin C.~Morey$^3$, Gabor Kunstatter$^4$,}
\end{center}

\centerline{\small \it $^1$ Natural Sciences Department, Metropolitan State University, Saint Paul, Minnesota, USA 55106}
\vskip 0 cm
\centerline{} 

\centerline{\small \it $^2$ Mathematics and Statistics Department, Metropolitan State University, Saint Paul, Minnesota, USA 55106}
\vskip 0 cm
\centerline{} 

\centerline{\small \it $^3$ School of Mathematics, University of Minnesota, Minneapolis, Minnesota, USA 55455}
\vskip 0 cm
\centerline{} 

\centerline{\small \it $^4$ Physics Department, University of Winnipeg, Winnipeg, MB Canada R3B 2E9}
\vskip 0 cm
\centerline{} 

\vspace{1cm}

\begin{abstract}
	We investigate the massless scalar ﬁeld perturbations, including the quasinormal mode spectrum and the ringdown waveform, of a regular black hole spacetime that was derived via the Loop Quantum Gravity inspired polymer quantization of spherical $4$D black holes.
	In contrast to most, if not all, of the other regular black holes considered in the literature, the resulting nonsingular spacetime has a single bifurcative horizon and hence no mass inflation. In the interior, the areal radius decreases to a minimum given by the Polymerization constant, $k$, and then re-expands into a Kantowski-Sachs universe.   We find indications that this black hole model is stable against small scalar perturbations.  We also show that an increase in the magnitude of $k$ will decrease the height of the quasinormal mode potential and give oscillations with lower frequency and less damping. 
\end{abstract}


\section{Introduction}

The black hole singularity is an open problem in theoretical physics.  The general consensus is that the singularity is the result of a classical treatment of spacetime and will not be present when quantum effects are considered.  Therefore, after the first specific proposal for a regular (singularity-free) black hole was presented by Bardeen \cite{reg1}, many regular black hole models have been proposed by various authors.  See  \cite{PoissonIsrael, reg2,reg3,reg4, reg5, reg6,reg7, reg9,reg10,reg11, reg12, reg14} for a partial list of the proposed regular black holes that mimic the behavior of the four dimensional Schwarzschild black hole at large distances.

In this paper, we are interested in studying how regularity of a black hole affects its response to perturbations, including the quasinormal mode (QNM) spectrum and the ringdown waveform. In addition to being, in principle, relevant to gravitational wave observations,  our calculations of the QNM spectrum are also important in determining the stability\cite{RW} of the black hole spacetime under consideration.

Regular black holes are usually constructed in a way that ensures the spacetime outside the horizon  of a macroscopic black hole is more or less indistinguishable from the classical black hole in General Relativity.  The metrics invariably contain a new parameter of dimension length that is usually taken to be the Planck length. Any departures from General Relativity are only detectable when the black hole radius is comparable to this length scale.  This does not deter us from investigating the ringdown wave signal of regular black holes because, in the absence of a quantum theory of gravity, it is not clear how big the new length scale in the regular black hole spacetime will be.  In addition, a regular black hole could potentially emerge from the effective model of yet unknown physics in which the new length scale is significantly larger than expected.  A relevant example is the regular black hole proposed in \cite{reg9}, in which the singularity resolution is caused by the brane-world scenario in \cite{RS}, where the new length scale can accept large values.  See also \cite{Teukolsky} where the authors choose the extra dimensionful parameter that appears in the dynamical Chern-Simons gravity to be very large compared to the Planck scale.


Most regular black holes have two horizons and  a conformal structure (Penrose diagram) that is similar to that of the Reissner-Nordstr\"{o}m black hole, with the new parameter playing the role of charge. The absence of a singularity in these cases is accompanied by a change in topology in the interior, but the problem of mass inflation \cite{MassInflation} along the inner horizon persists. 

In this paper we consider a different type of  regular black hole proposed by Peltola and one of us in \cite{pk09}.  The line element of this black hole is given in Eq.\ (\ref{PK-LE}), where $k$ is a new (possibly quantum) length scale.  When $k=0$ the Schwarzschild spacetime is recovered. This regular black hole differs from most, if not all, of the other proposed models in the literature in at least one important respect. It has only a single horizon, which  eliminates the problem of mass inflation. The single horizon also means that the global structure is quite different:   the Penrose diagram describes at each time slice an Einstein-Rosen bridge just as in the Schwarzschild solution. The difference is that the radius of the throat shrinks to a minimum determined by the parameter $k$ before re-expanding  to an infinite Kantowski-Sachs spacetime, realizing the proposed scenarios for  universe generation within black hole horizons\cite{Frolov, Brandenberger}.
The single horizon also means that the analytic structure of the solution is quite different.  In this paper, we explore whether  the difference in the analytic structure leads to a difference in the QNM spectrum. 


The regular black hole that we consider was derived in \cite{pk09} using a method inspired by Loop Quantum Gravity (LQG).  LQG starts with the classical phase space for the gravitational field, which consists of an SU(2) valued connection $A^i_a(t,\vec r)$ and its canonically conjugate momentum $E_i^a(t,\vec r)$, where $i=1,2,3$ represents spatial indices and $a=1,2,3$ represents SU(2) indices.  A fundamental feature of LQG is that the observable corresponding to area has a discrete spectrum. 

Since a full quantum treatment in LQG is very complex even in the case of spherical symmetry, most work uses the so-called effective field theory approach. This effective theory is derived by first constructing the quantum theory using polymer quantization \cite{afw,halvorson}  in which at least one of the canonical variables takes its value on the real line with discrete topology. Just as in full LQG, this necessarily introduces a new length scale.  The effective theory is obtained by taking  the classical ($\hbar\to 0$) limit of the polymer quantum theory while keeping the discrete length scale fixed. The result is a set of effective equations for the geometrical variables that can be solved to obtain the metric. 

Spherically symmetric gravity necessarily has a Killing vector that is timelike exterior to the horizon, so one  considers the black hole interior where the Killing vector is spacelike and spatial slices are homogeneous.  The classical solutions have non-trivial dynamics in the interior and after a suitable choice of coordinates one is left with a finite dimensional quantum mechanical system to solve. 

As in earlier works\cite{ashtekar05, modesto06, boehmer07, pullin08}, the authors of \cite{pk09} applied the  effective field theory approach to the symmetry reduced black hole Hamiltonian written in terms of the two configuration space variables: the areal radius\footnote{The areal radius  $r=\sqrt{\frac{A}{4\pi}}$ where $A$ is the area of the sphere generated by the action of the angular Killing vectors starting from any point} $r$ and a function $\rho$ obtained from the conformal mode of the metric. Only $r$ is invariant under all coordinate transformations that preserve spherical symmetry so that after gauge fixing it remains as the single physical degree of freedom. {One} key difference between \cite{pk09} and earlier methods is that polymer quantization is applied only to $r$ in order to take into account the discreteness of area in LQG. The other variable $\rho$ is assumed continuous and  quantized using the standard Schr\"odinger representation. This simplifying approximation is justified by noting that in the full LQG treatment the discretization scale  associated with $\rho$ is considerably smaller than that of $r$.
The effective field equations that emerge are readily solvable and lead to  the following line element for the black hole exterior:
\begin{equation}
	ds^2=-\left(\sqrt{1-\frac{k^2}{r^2}}-\frac{2GM}{r}\right)dt^2+\frac{dr^2}{\left(\sqrt{1-\frac{k^2}{r^2}}-\frac{2GM}{r}\right)\left(1-\frac{k^2}{r^2}\right)}+r^2 d\Omega^2~, 
	\label{PK-LE}
\end{equation}
where $t$ is the Schwarzschild time, $r$ is  the areal radius and $k$ is the polymer length scale. 
Eq.~(\ref{PK-LE}) describes the black hole whose scalar perturbations we consider.



The perturbations of different regular black hole models are investigated in, for example, \cite{QNMLQGbh0}-\cite{QNMreg15}.  The high overtone QNMs for massless scalar perturbations of the regular black hole model (\ref{PK-LE}) are analytically calculated by Babb and two of us in \cite{Babb}.  In this paper, we investigate the perturbations, including the low overtone QNMs and the ringdown waveform, of a massless scalar field coupled to the regular black hole background defined by the line element (\ref{PK-LE}).  

We structure the paper as follows. In Sec.\ \ref{Sec:WE}, we set up the problem by introducing the QNM wave equation. In Sec.\ \ref{Sec:WKBQNM}, we calculate the QNM complex frequencies using the $6^{th}$ order WKB method.  In Sec.\ \ref{Sec:IAIM}, we calculate the QNMs using the Asymptotic Iteration Method and compare the results to those in Sec.\ \ref{Sec:WKBQNM}.  In Sec.\ \ref{Sec:ringdown}, we produce and analyze the ringdown waveforms for various values of the length scale $k$.  Finally, the summary and conclusion are presented in Sec.\ \ref{Sec:conclusions}.

\section{Wave Equation}
\label{Sec:WE}

A massive scalar field in curved spacetime obeys the Klein-Gordon equation
\begin{equation}
\frac{1}{\sqrt{-g}}\partial_\mu\left( {\sqrt{-g}}g^{{\mu}{\nu}}\partial_\nu{\Phi} \right)-m^2 \Phi=0~,
\label{KG-wave-eq}
\end{equation}
where $m$ is the mass of the scalar field, $g_{\mu\nu}$ is the metric and $g$ is its determinant. Here we use units $c=\hbar=1$.  In a completely general, spherically symmetric, static spacetime, the line element can be written as
\begin{equation}
ds^2=-A(r)dt^2+B(r)^{-1}dr^2+r^2d\Omega^2~.
\label{spherical-le}
\end{equation} 
We can separate variables by writing
\begin{equation}
\Phi(t,r, \theta, \phi) = Y_l(\theta, \phi)\Psi(t,r)/r~,
\label{seperate-variable}
\end{equation}
where $Y_l(\theta, \phi)$ are spherical harmonics with the multipole number $l=0,1,2,\dots$.  Combining Eqs.\ (\ref{KG-wave-eq}), (\ref{spherical-le}) and (\ref{seperate-variable}) gives
\beeq
\frac{\partial^2\Psi}{\partial t^2}+\left(-\frac{\partial^2}{\partial r_*^2}+V(r)\right)\Psi=0~,
\label{WE-time}
\eneq  
where $r_*$ is the tortoise coordinate linked to the radial coordinate according to
\beeq
dr_*=\frac{dr}{\sqrt{A(r)B(r)}}~,
\label{tortoise}
\eneq
and 
\beeq
V(r)=A(r) \left[\frac{l(l+1)}{r^2}+m^2\right]+ \frac{1}{2r} \frac{d}{dr} \left[ A(r)B(r) \right]~.
\label{scalarV}
\eneq
If we assume the perturbations depend on time as 
\beeq
\Psi(t,r)=e^{-i \omega t} \psi(r)~,
\label{TimeDepend}
\eneq  
we can use Eq.~(\ref{WE-time}) to obtain
\beeq
\frac{d^2 \psi}{dr_*^2}+\left[\omega^2-V(r)\right]\psi=0~,
\label{WEnoTime}
\eneq  
where $\omega$ is the complex QNM frequency to be determined.

For the specific line element (\ref{PK-LE}), the potential (\ref{scalarV}) takes the following form
\begin{eqnarray}
V(r) &=& 
\left(\sqrt{1-\frac{k^2}{r^2}}-\frac{2GM}{r}\right) \nonumber \\
&& \times \left[\frac{l(l+1)}{r^2}+m^2  +\frac{2k^2}{r^4} \left(\sqrt{1-\frac{k^2}{r^2}}-\frac{2GM}{r}\right) +\frac{2GM}{r^3} \right]~.
\label{Vqnm}
\end{eqnarray}
The tortoise coordinate for this spacetime can be derived by combining Eqs.~(\ref{PK-LE}) and (\ref{tortoise}):
\begin{eqnarray}
r_* &=& r+\sqrt{(2GM)^2+k^2}\ln\left(\sqrt{r^2-k^2}-2GM\right)+2GM \ln\left(r+\sqrt{r^2-k^2}\right) \nonumber \\
&& -\sqrt{(2GM)^2+k^2}\ln\left(\frac{\sqrt{(2GM)^2+k^2}~r+2GM\sqrt{r^2-k^2}+k^2}{2GM}\right) + C ~,
\label{r-tortoise}
\end{eqnarray}
where $C$ is the constant of integration.  In the rest of this paper, we choose $C$ so that the peak of the QNM potential (\ref{Vqnm}) in the tortoise coordinate is centered at $r_*=0$.  Note that when $k=0$, the relationship between the tortoise coordinate and the radial coordinate reduces to the Schwarzschild case where $r_*=r+2GM \ln(r-2GM)+C$.

With our choice of time-dependence (\ref{TimeDepend}), the boundary conditions at the event horizon and infinity are, respectively,
\beeq
\begin{array}{ll}
	\psi(x) \rightarrow e^{-i\omega r_*}  & \mbox{as $r_* \rightarrow -\infty$ ($r \rightarrow \sqrt{(2GM)^2+k^2}$)}~,\\
	\psi(x) \rightarrow e^{i\omega r_*}  & \mbox{as $r_* \rightarrow \infty$ ($r \rightarrow \infty$)}~.
\end{array}       
\label{B.C.}
\eneq

\section{The WKB Method}
\label{Sec:WKBQNM}

To calculate the QNMs of the regular black hole spacetime (\ref{PK-LE}), we first use the WKB method.  This method was originally applied to the problem of black hole perturbations by Schutz and Will in \cite{WKB1}.  The formula for the $3^{rd}$ order WKB method is derived in \cite{WKB2} and it is extended to the $6^{th}$ order by Konoplya in \cite{WKB-Konoplya}.  

It is shown in \cite{WKB2} that in order to determine the QNMs using the WKB method, one needs to solve the following equation
\beeq
\frac{i \left[ \omega^2 -V(r_{*})|_{\bar{r}_*}\right]}{\sqrt{2V''(r_*)|_{\bar{r}_*}}} -\overset{N}{\underset{j=2}{\sum}} \Lambda_j(n)=n+\dfrac{1}{2},
\label{WKBorder}
\eneq
where $\bar{r}_*$ is the location of the maximum of the QNM potential $V(r_*)$ in the tortoise coordinate. Prime indicates differentiation with respect to $r_*$,  and $\Lambda_j(n)$ are the WKB correction terms. $N$ indicates the order of the WKB method.  $\Lambda_{2,3}$ are given in \cite{WKB2}\footnote{$\Lambda_2$ in \cite{WKB2} is missing a factor of $i$ in the numerator.} and $\Lambda_{4,5,6}$ can be found in \cite{WKB-Konoplya}.

In Table I, we provide the QNM complex frequencies for different values of the root number $n$ and the multipole number $l$.  As is indicated in \cite{WKB-Konoplya}, the WKB method works more accurately for lower values of $n$ and higher values of $l$.   For example, for $l=0,1,2$ the table shows the only reliable roots we can find, while for $l=6$ we find many more roots but include only eight of them for brevity.  

\begin{figure}[th!]
	\begin{center}
		\includegraphics[height=3.9cm]{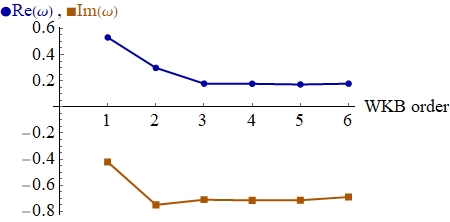}
		\includegraphics[height=3.9cm]{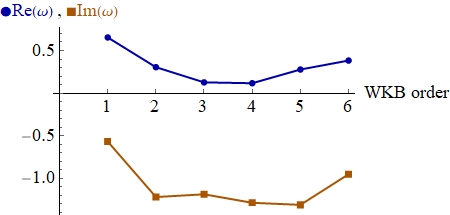}
		\includegraphics[height=3.9cm]{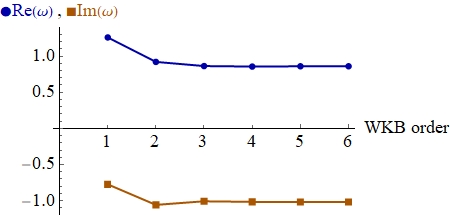}
	\end{center}
	\vspace{-0.7cm}
	\caption{\footnotesize The real (top in blue) and imaginary (bottom in brown) parts of $2GM \omega$ as a function of the order of the WKB method for $\frac{k}{2GM}=0.5$.  On the top left $l = 0$ and $n = 1$. On the top right $l=0$ and $n=2$. On the lower plot $l=2$ and $n=2$.\cite{KonoplyaCode}}
	\label{w-WKBorder}
\end{figure}

The reliability of the roots is determined by comparing the results for different orders of the WKB method.   For example, in Figure \ref{w-WKBorder} we see the real and imaginary parts of $\omega$ converge as we increase the order of the WKB method for $l=0$ and $n=1$.  That is not the case for  $l=0$ and $n=2$. Convergence for higher values of $n$ improves with higher values of $l$.  We illustrate this in Figure \ref{w-WKBorder}, where we plot the real and imaginary parts of $\omega$ for $l = 2$ and $n = 2$ to be compared with the case of $l=0$ and $n=2$.

We also look at convergence plots, similar to Figure \ref{w-WKBorder}, to determine if the accuracy of the WKB method depends on the value of the parameter $k$.  We find no correlation.

\vspace{0.5cm}
\footnotesize
\begin{tabular}{cccccc}
	\multicolumn{6}{c}{Table I:  $2GM \omega$ for different values of $\frac{k}{2GM}$ using $6^{th}$ order WKB method\cite{KonoplyaCode}} \\ 
	\hline
	$n,l$ & $\frac{k}{2GM}=0$ &  $\frac{k}{2GM}=0.01$ & $\frac{k}{2GM}=0.1$  &  $\frac{k}{2GM}=0.5$  &\\ 
	\hline 
	0,0 & $0.220928 - 0.201638i$ &  $0.220914 - 0.201635i$ & $0.220132 - 0.200850 i$ & $0.205181 - 0.185673i$ \\ 
	1,0 & $0.178046 - 0.689103i$ &  $0.178030 - 0.689114i$ & $0.177375 - 0.686627i$ & $0.165269 - 0.630781i$  \\ 
	\hline 
	0,1 & $0.585819 - 0.195523i$ &  $0.585799 - 0.195517i$ & $0.583829 - 0.194900  i$ & $0.542441 - 0.181808i$ \\ 
	1,1 & $0.528942 - 0.613037i$ &  $0.528925 - 0.613017i$ & $0.527155 - 0.611100 i$ & $0.489571 - 0.569794i$  \\ 
	2,1 & $0.462028 - 1.08433i$ &  $0.462015 - 1.08429i$ & $0.460532 - 1.08090i$ &$0.427411 - 1.00681i$ \\
	\hline 
	0,2 & $0.967284 - 0.193532i$ &  $0.967251 - 0.193526i$ & $0.964057 - 0.192895  i$ & $0.896660 - 0.179678i$ \\ 
	1,2 & $0.927693 - 0.591254i$ &  $0.927662 - 0.591234i$ & $0.924594 - 0.589312i$ & $0.859622 - 0.548953 i$  \\ 
	2,2 & $0.860771 - 1.01740i$ &  $0.860743 - 1.01737i$ & $0.857891 - 1.01407i$ &$0.796985 - 0.944587 i$ \\
	3,2 & $0.786413 - 1.47977i$ &  $0.786387 - 1.47972 i$ & $0.783779 - 1.47493i$ &$0.727102 - 1.37375i$ \\
	4,2 & $0.725174 - 1.97659i$ &  $0.725151 - 1.97653 i$ & $0.722738 - 1.97016i$ &$0.668827 - 1.83492i$ \\
    \hline 
	0,6 & $2.50377 - 0.192610i$ &  $2.50369 - 0.192604i$ & $2.49547 - 0.191973 i$ & $2.32199 - 0.178751i$ \\ 
	1,6 & $2.48750 - 0.579473i$ &  $2.48742 - 0.579453i$ & $2.47926 - 0.577555i$ & $2.30680 - 0.537770 i$  \\ 
	2,6 & $2.45569 - 0.971202i$ &  $2.45561 - 0.971170i$ & $2.44754 - 0.967988 i$ &$2.27710 - 0.901286 i$ \\
	3,6 & $2.40978 - 1.37082i$ &  $2.40972 - 1.37077 i$ & $2.40178 - 1.36628i$ &$2.23422 - 1.27209i$ \\
	4,6 & $2.35191 - 1.78094i$ &  $2.35184 - 1.78088 i$ & $2.34411 - 1.77505 i$ &$2.18014 - 1.65259 i$ \\
	5,6 & $2.28477 - 2.20360i$ &  $2.28469 - 2.20352i$ & $2.27718 - 2.19631i$ &$2.11734 - 2.04467 i$ \\
	6,6 & $2.21133 - 2.64016i$ &  $2.21126 - 2.64008i$ & $2.20399 - 2.63143i$ &$2.048570 - 2.44960 i$ \\
	7,6 & $2.13469 - 3.09138i$ &  $2.13462 - 3.09128 i$ & $2.12759 - 3.08115 i$ &$1.97669 - 2.86804 i$ \\
	\label{Table1}
\end{tabular} 
\normalsize

\newpage 

\section{Improved Asymptotic Iteration Method}
\label{Sec:IAIM}

As mentioned earlier, the WKB method becomes less accurate for lower values of $l$ and higher values of $n$.  
Since the accuracy of the WKB method is in question, it is useful to compare the results to those found using other methods.

In this section, we use the improved Asymptotic Iteration Method (AIM) described in \cite{AIM} to determine QNM frequencies.  We briefly explain the AIM below and show how to adapt the method to our black hole model.

Given a differential equation of the form
\begin{equation}
\chi'' = \lambda_0(x) \chi' + s_0(x) \chi ~,
\label{AIMequation}
\end{equation}
one can express higher derivatives of $\chi$ in terms of $\chi'$ and $\chi$ as follows:
\begin{equation}
\chi^{(n+1)} = \lambda_{n-1}(x) \chi' + s_{n-1}(x) \chi ~,
\end{equation}
where
\begin{gather}
\lambda_n(x) = \lambda_{n-1}'(x) + s_{n-1}(x) + \lambda_0(x)\lambda_{n-1}(x) \nonumber \\
s_n(x) = s_{n-1}'(x) + s_0(x)\lambda_{n-1}(x)~. 
\end{gather}
We now expand $\lambda_n$ and $s_n$ in a Taylor series around some point $x_0$:
\begin{gather}
\lambda_n(x) = \sum_{i=0}^{\infty} c_n^i(x-x_0)^i \nonumber\\
s_n(x) = \sum_{i=0}^{\infty} d_n^i(x-x_0)^i ~,
\end{gather}
which allows us to rewrite the recurrence relations for $\lambda_n$ and $s_n$ in terms of the coefficients $c_n$ and $d_n$:
\begin{gather}
c_n^i = (i+1)c_{n-1}^{i+1} + d_{n-1}^i + \sum_{k=0}^i c_0^k c_{n-1}^{i-k}  \\
d_n^i = (i+1)c_{n-1}^{i+1} + \sum_{k=0}^i d_0^k c_{n-1}^{i-k} ~.
\end{gather}
We then make the assumption that, for large $n$, 
\begin{equation}
\frac{s_n(x)}{\lambda_n(x)} = \frac{s_{n-1}(x)}{\lambda_{n-1}(x)} ~,
\end{equation}
which leads to the equation
\begin{equation}
s_n\lambda_{n-1}-s_{n-1}\lambda_n=0 ~.
\end{equation}
In terms of the Taylor series coefficients, the above equation is equivalent to 
\begin{equation}
d_n^0 c_{n-1}^0 - d_{n-1}^0 c_n^0 = 0~,
\label{recurrenceforcoeffs}
\end{equation}
which gives an equation in terms of the QNM frequencies.

To apply AIM, we need to rewrite our wave equation (\ref{WEnoTime}) in the form (\ref{AIMequation}). First, we find it more convenient to work in the coordinate
\beeq
x = \sqrt{r^2-k^2}~.
\label{new-coordinate}
\eneq
The above coordinate is also used in \cite{Babb} to calculate the high overtone QNMs of the regular black hole (\ref{PK-LE}).  

In the new coordinate, the wave equation is
\beeq
\begin{array}{ll}
	x (x-2GM)\frac{d^2 \psi}{dx^2}+\left(x-\frac{x^2(x-2GM)}{x^2+k^2}  \right)\frac{d \psi}{dx} \\ 
	-\left\{\frac{\rho^2 x (x^2+k^2)}{x-2GM}+ \frac{l(l+1) x }{\sqrt{x^2+k^2}}+m^2 x {\sqrt{x^2+k^2}}  + \frac{2k^2 x (x - 2GM)}{(x^2+k^2)^2} +  \frac{2GM x}{x^2+k^2} \right\}\psi=0~,
\end{array}
\label{WE-X}
\eneq  
where $\rho=-i \omega$.  It is easy to show that the above equation reduces to the wave equation of a massive scalar field coupled to the Schwarzschild background when $k=0$.

We can now combine equations (\ref{r-tortoise}), (\ref{B.C.}), and (\ref{new-coordinate}) to find the asymptotic behavior of the wavefunction:
\beeq
\psi \overset{x \rightarrow 2GM}{\larrow}  (x-2GM)^{\rho \sqrt{(2GM)^2+k^2}}~~\mbox{and}~~\psi \overset{x \rightarrow \infty}{\larrow}  x^{-2GM\rho} e^{-\rho x}~.
\label{BC-X}
\eneq
We then write a solution to the wave equation (\ref{WE-X}) that has the correct asymptotic behavior at the boundaries in the following form:
\beeq
\psi(x) =  \left( \frac{x-2GM}{x} \right)^{\rho \sqrt{(2GM)^2+k^2}}  \left(\frac{x}{2GM}\right)^{-2GM\rho} e^{-\rho (x-2GM)} \chi(x).
\label{WaveF-X}
\eneq

To follow the same steps taken in \cite{AIM}, we introduce the change of variable
\begin{equation}
\xi = 1- \frac{2GM}{x}~.
\label{}
\end{equation}
Substituting this into (\ref{WE-X}) leads to a differential equation for $\chi(\xi)$ in the form (\ref{AIMequation}) where 
\begin{equation}
\lambda_0 = -\frac{k^2(\xi-1)^2(2\xi-1)+(2GM)^2(3\xi-1)}{((2GM)^2 + k^2(\xi-1)^2)(\xi-1)\xi} + \frac{4GM\rho}{(\xi-1)^2}-\frac{4GM\rho}{\xi-1}-\frac{2\rho \sqrt{k^2+(2GM)^2}}{\xi} 
\end{equation}
and
\begin{eqnarray}
s_0 &=& -\frac{2 GM}{\xi^2(\xi-1)^4}\left( \frac{2 GM \xi(\xi-1)^3 }{k^2 (\xi-1)^2 + (2GM)^2}
+ \frac{\xi (\xi-1)^2 l(l+1)}{\sqrt{k^2 (\xi-1)^2 + (2GM)^2}} - 2GM\rho^2 \right) \nonumber \\
&& + k^2 \left(\frac{2(2GM)^2}{\left(k^2
	(\xi-1)^2+(2GM)^2\right)^2} 
+ \frac{\rho ^2}{\xi^2 (\xi-1)^2}\right)-\frac{4 GM \rho ^2 \sqrt{k^2+(2GM)^2}}{\xi(\xi-1)} \nonumber \\
&&	- \frac{\rho  \left((\xi-1)^2 \sqrt{k^2+(2GM)^2}+2 GM
	(\xi-2) \xi \right) \Big( k^2 (\xi-1)^2 (2 \xi-1)+(2GM)^2 (3
	\xi-1)\Big)}{(\xi-1)^3 \xi^2 \left(k^2 (\xi-1)^2+(2GM)^2 \right)} \nonumber \\
&& + \frac{\rho  \left(\sqrt{k^2+(2GM)^2}-\rho  \left(k^2+(2GM)^2\right)\right)}{\xi^2} 
       +\frac{4 GM \rho ^2 \sqrt{k^2+(2GM)^2}}{(\xi-1)^2 \xi}\nonumber \\
&& + \frac{(2GM)^2 \rho ^2}{(\xi-1)^3}-\frac{(2GM)^2 \rho
	^2}{(\xi-1)^4}+\frac{2 GM \rho  (1-2 GM \rho )}{(\xi-1)^2}-\frac{2 GM
	\rho  (2-2GM \rho)}{(\xi-1)^3} ~.
\end{eqnarray}

We now expand $\lambda_0$ and $s_0$ in a Taylor series around a point $\xi_0$.  After selecting an appropriate depth ($n$ in Eq.\ (\ref{recurrenceforcoeffs})), we substitute the coefficients into ($\ref{recurrenceforcoeffs}$) to obtain an equation in $\rho$.  A root finder is then used to find the QNMs.  Although the choice of $\xi_0$ should not make a difference, in practice there are small variations when changing $\xi_0$.  We find setting $\xi_0$ to the $\xi$-value corresponding to the  maximum of the potential produces the best results.

In the following table, for comparison, we present the QNMs for the Schwarzschild black hole obtained by Leaver's  Continued Fraction Method\cite{Leaver} (CFM) and AIM. To produce the roots using Leaver's method, we wrote a program in {\em Mathematica} in which we incorporated Nollert's improvement\cite{Nollert1}.  As one can see, the results of the two methods are in good agreement for lower values of overtone number $n$ and higher values of multipole number $l$.

\begin{center}
\vspace{0.5cm}
\footnotesize
\begin{tabular}{cccc} 
	\multicolumn{4}{c}{Table II: $2GM \omega$ for $l=0$ and  $l=1$} for Schwarzschild\\ 
	\multicolumn{4}{c}{black hole using CFM and AIM} \\
	\hline
	$n,l$ & CFM &  AIM  \\ 
	\hline 
	0,0 &$0.220910 - 0.209791i$ &  $0.220910-0.209791 i$  \\ 
	1,0 &$0.172234 - 0.696105i$ &  $0.172416-0.696149 i$   \\ 
	2,0 & $0.151484 - 1.20216i$ &  $0.151053-1.20040i$  \\ 
	\hline 
	0,1 & $0.585872-0.195320  i$ & $0.585872-0.195320i$  \\ 
	1,1 & $0.528897-0.612515  i$ & $0.528897-0.612515i$  \\ 
	2,1 &$0.459079-1.08027   i$ & $0.459079-1.08027i$ \\
	3,1 &$0.406517-1.57660   i$ & $0.406510-1.57659i$ \\
	\hline 
	0,2 & $0.967290-0.193518i$ &  $0.967288-0.193518i$  \\ 
	1,2 & $0.927701-0.591208 i$ &  $0.927701-0.591208i$   \\ 
	2,2 & $0.861088-1.01712 i$ &  $0.861088-1.01712i$  \\
	3,2 & $0.787726-1.47619i$ &  $0.787726-1.47619i$  \\
	4,2 & $0.722598-1.95984i$ &  $0.722598-1.95984i$  \\
	5,2 & $0.669799 - 2.45682i$ &  $0.669800-2.45682i$  \\
	6,2 & $0.627772 - 2.95991i$ &  $0.627775-2.95990i$  \\
	\label{Table1}
\end{tabular} 
\normalsize
\end{center}

Finally, we present the results for different values of the parameter $k$ in the table below.  Note that we found more roots for $l=2$ and $l=6$, but we only include seven roots for brevity.  Additional roots are included in the graphs.

\vspace{0.5cm}
\footnotesize
\begin{tabular}{cccccc}
	\multicolumn{6}{c}{Table III:  $2GM \omega$ for different values of $k/2GM$ using AIM} \\ 
	\hline
	$n,l$ & $\frac{k}{2GM}=0$ &  $\frac{k}{2GM}=0.01$ & $\frac{k}{2GM}=0.1$  &  $\frac{k}{2GM}=0.5$  &\\ 
	\hline 
	0,0 & $0.22091-0.20979i$ &  $0.22090-0.20978i$ & $0.22005-0.20907 i$ & $0.20218-0.19421i$ \\ 
	1,0 & $0.17242-0.69615i$ &  $0.17241-0.69612i$ & $0.17169-0.69378i$ & $0.15478-0.64486i$  \\ 
	2,0 & $0.15105-1.2004i$ &  $0.15104-1.2004i$ & $0.15025-1.1964i$ & $0.12868-1.1147i$  \\ 
	\hline 
	0,1 & $0.58587-0.19532i$ &  $0.58585-0.19531i$ & $0.58389-0.19467i$ & $0.54253-0.18131i$ \\ 
	1,1 & $0.52890-0.61251i$ &  $0.52888-0.61249i$ & $0.52709-0.61049i$ & $0.48898-0.56852i$  \\ 
	2,1 & $0.45908-1.0803i$ &  $0.45906-1.0802i$ & $0.45749-1.0767i$ &$0.42280-1.0025i$ \\
	3,1 & $0.40651-1.57656 i$ &  $0.27463-5.1071 i$ & $0.40508-1.5714i$ &$0.37180-1.4628 i$ \\
	\hline 
	0,2 & $0.96729-0.19352i$ &  $0.96726-0.19351i$ & $0.96406-0.19288  i$ & $0.89667-0.17963i$ \\ 
	1,2 & $0.92770-0.59121i$ &  $0.92767-0.59119i$ & $0.92460-0.58926i$ & $0.85961-0.54875 i$  \\ 
	2,2 & $0.86109-1.0171i$ &  $0.86106-1.0171i$ & $0.85820-1.0138i$ & $0.79711-0.94396 i$ \\
	3,2 & $0.78773-1.4762i$ &  $0.78770-1.4761 i$ & $0.78507-1.4713i$ &$0.72794-1.3698i$ \\
	4,2 & $0.72260-1.9598i$ &  $0.72257-1.9598 i$ & $0.72014-1.9534i$ &$0.66593-1.8183i$ \\
	5,2 & $0.66980-2.4568i$ &  $0.66978-2.4567 i$ & $0.66750-2.4487i$ &$0.61483-2.2790i$ \\
	6,2 & $0.62778-2.9599i$ &  $0.62775-2.9598i$ & $0.62559-2.9502i$ &$0.57312-2.7453i$ \\
	\hline 
	0,6 & $2.5038-0.19261i$ &  $2.5037-0.19260i$ & $2.4955-0.19197 i$ & $2.3220-0.17875i$ \\ 
	1,6 & $2.4875-0.57947i$ &  $2.4874-0.57945i$ & $2.4793-0.57755i$ & $2.3068-0.53777 i$  \\ 
	2,6 & $2.4557-0.97120 i$ &  $2.4556-0.97117i$ & $2.4475-0.96799 i$ &$2.2771-0.90129 i$ \\
	3,6 & $2.4098-1.3708i$ &  $2.4097-1.3708 i$ & $2.4018-1.3663i$ &$2.2342-1.2721i$ \\
	4,6 & $2.3521-1.7810i$ &  $2.3520-1.7809 i$ & $2.3443-1.7751 i$ &$2.1803-1.6526 i$ \\
	5,6 & $2.2854-2.2035i$ &  $2.2853-2.2035i$ & $2.2778-2.1963i$ &$2.1179-2.0446 i$ \\
	6,6 & $2.2130-2.6396i$ &  $2.2129-2.6395i$ & $2.2057-2.6308i$ &$2.0501-2.4490 i$ \\
	\label{Table1}
\end{tabular} 
\normalsize

\begin{figure}[th!]
	\begin{center}
		\includegraphics[height=8.cm]{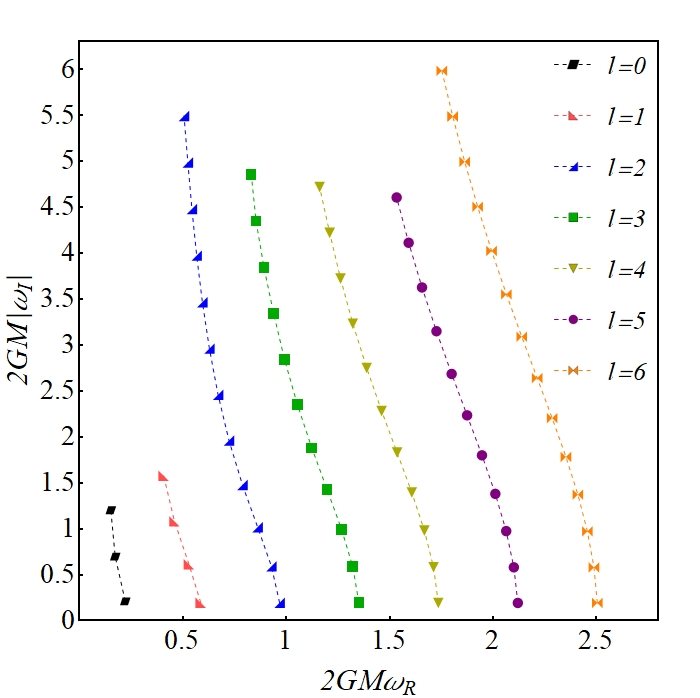}
		\includegraphics[height=8.cm]{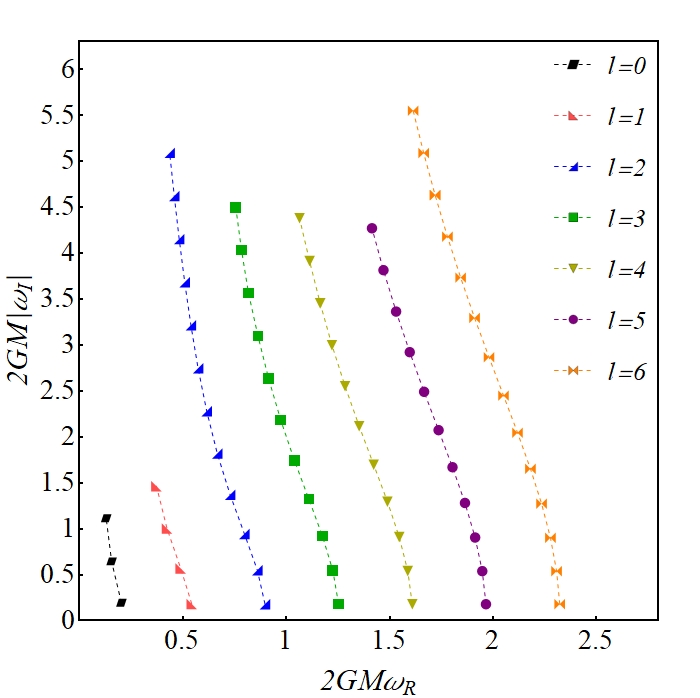}
	\end{center}
	\vspace{-0.7cm}
	\caption{\footnotesize Scalar QNM spectrum for  $\frac{k}{2GM}=0$ (Schwarzschild), on the left, and $\frac{k}{2GM}=0.5$, on the right, for different values of $l$.}
	\label{AIM-different-l}
\end{figure}

\begin{figure}[th!]
	\begin{center}
		\includegraphics[height=8.cm]{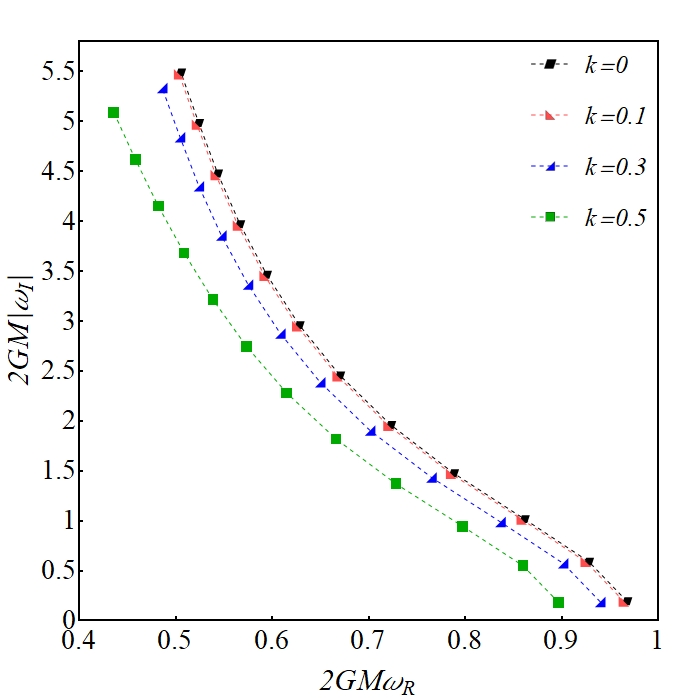}
	\end{center}
	\vspace{-0.7cm}
	\caption{\footnotesize Scalar QNM spectrum for $l=2$ for different values of  $k$ given in the length unit of $2GM$. }
	\label{AIM-different-k}
\end{figure}

In Figure \ref{AIM-different-l}, we plot the QNM spectrum for $\frac{k}{2GM}=0$ (Schwarzschild)  and $\frac{k}{2GM}=0.5$ for different values of $l$.  As $l$ increases, the real part of the QNM frequency increases, by a nearly fixed amount, and the imaginary part decreases. As $k$ increases, the magnitudes of the real and imaginary parts of $\omega$ decrease.  This becomes more evident in Figure \ref{AIM-different-k}, where we show the QNMs for $l=2$ for different values of $k$.

The roots produced by AIM become more accurate as the depth ($n$ in Eq.\ (\ref{recurrenceforcoeffs})) increases.  The roots included in Tables II/III were calculated using a depth of $110$.  However, to determine which of the roots ($\omega_n$) found at this depth are most reliable, we compare them with roots calculated at a depth of 105 ($\omega'_n$). All the roots in Tables II/III have at least $2GM|\omega_n-\omega'_n|<0.0034$.  For higher values of $l$, AIM converges more quickly.  For example, all of the roots in Table III for $l=6$ have $2GM|\omega_n-\omega'_n|<1.8 \times 10^{-16}$.

\newpage

\section{Ringdown Waveform}
\label{Sec:ringdown}

In this section, the units are chosen such that $2GM=1$.  To generate the ringdown waveform, we numerically solve the time-dependent wave equation (\ref{WE-time}) using the initial data
\beeq
\Psi(r_*,0)={\cal A} \exp \left(- \frac{(r_*-\bar{r}_{*})^2}{2\sigma^2} \right),~  \partial_t \Psi|_{t=0}=-\partial_{r_*} \Psi(r_*, 0)~,
\label{GaussianWave}
\eneq  
where we use $\sigma=1$, $\bar{r}_*=-40$, and ${\cal A}=20$.  We choose the observer to be located at $r_*=90$.   To carry out the calculations, we use the built-in {\em Mathematica} commands for solving partial differential equations.

\begin{figure}[th!]
	\begin{center}
		\includegraphics[height=5.2cm]{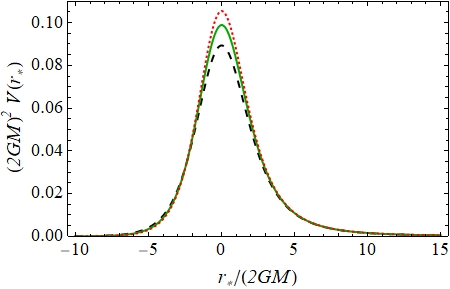}
		\includegraphics[height=5.2cm]{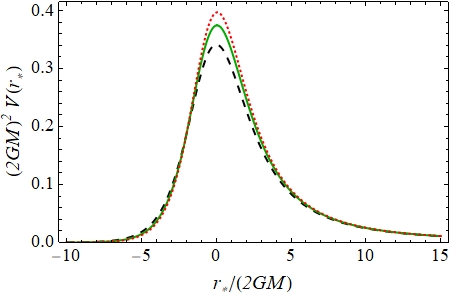}
	\end{center}
	\vspace{-0.7cm}
	\caption{\footnotesize Potential versus tortoise coordinate for $l=0$ (left) and $l=1$ (right).  In both graphs, for comparison, we include the cases $\frac{k}{2GM}=0$ in dotted red, $\frac{k}{2GM}=0.3$ in solid green, and $\frac{k}{2GM}=0.5$ in dashed black.}
	\label{potential}
\end{figure}

In Figure \ref{potential}, we compare the shape of the potential for $k=0$, $k=0.3$, and $k=0.5$.  As one can see, the height of the potential decreases as $k$ increases.  

\begin{figure}[th!]
	\begin{center}
		\includegraphics[height=5.2cm]{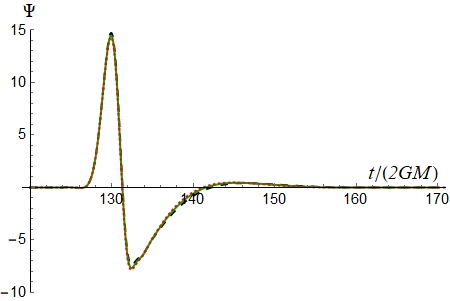}
		\includegraphics[height=5.2cm]{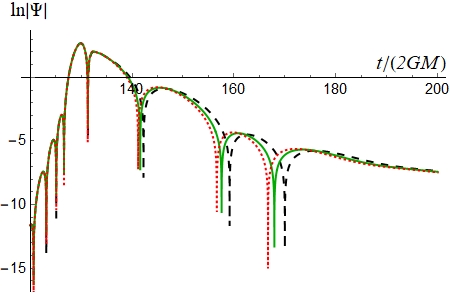}
	\end{center}
	\vspace{-0.7cm}
	\caption{\footnotesize $\Psi$ (left) and $\ln |\Psi|$ (right) as a function of time for $l=0$.  In both graphs, we include the cases $\frac{k}{2GM}=0$ in dotted red, $\frac{k}{2GM}=0.3$ in solid green, and $\frac{k}{2GM}=0.5$ in dashed black. }
	\label{l=0ringdown}
\end{figure}
\begin{figure}[th!]
	\begin{center}
		\includegraphics[height=5.2cm]{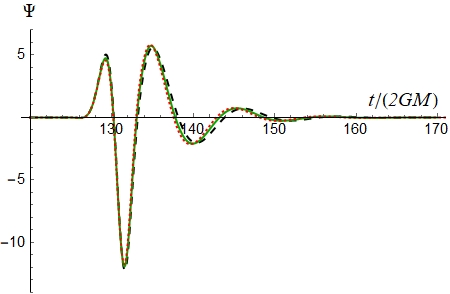}
		\includegraphics[height=5.2cm]{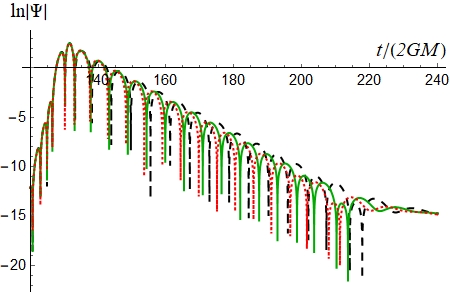}
	\end{center}
	\vspace{-0.7cm}
	\caption{\footnotesize $\Psi$ (left) and $\ln |\Psi|$ (right) as a function of time for $l=1$.  In both graphs, we include the cases $\frac{k}{2GM}=0$ in dotted red, $\frac{k}{2GM}=0.3$ in solid green, and $\frac{k}{2GM}=0.5$ in dashed black.}
	\label{l=1ringdown}
\end{figure}
\begin{figure}[th!]
	\begin{center}
		\includegraphics[height=5.2cm]{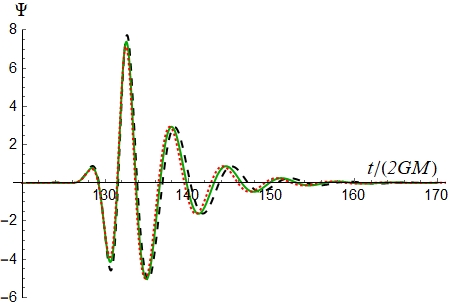}
		\includegraphics[height=5.2cm]{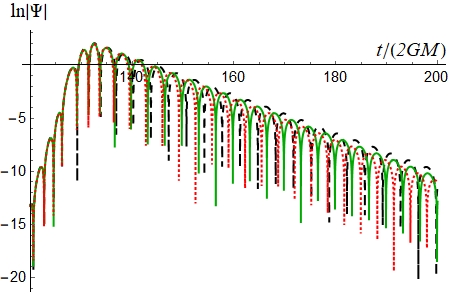}
	\end{center}
	\vspace{-0.7cm}
	\caption{\footnotesize$\Psi$ (left) and $\ln |\Psi|$ (right) as a function of time for $l=2$.  In both graphs, we include the cases $\frac{k}{2GM}=0$ in dotted red, $\frac{k}{2GM}=0.3$ in solid green, and $\frac{k}{2GM}=0.5$ in dashed black.  To better distinguish between the three cases, we stop at $\frac{t}{2GM}=210$ in the log graph.}
	\label{l=2ringdown}
\end{figure}

In Figures \ref{l=0ringdown}, \ref{l=1ringdown}, and \ref{l=2ringdown} we provide the ringdown waveform $\Psi$ (left)  and $\ln|\Psi|$ (right), as a function of time, for the potential (\ref{Vqnm}) with $l=0, 1, 2$ respectively.  We include the cases $k=0$, $k=0.3$, and $k=0.5$.  The oscillation periods are easier to see in the log plot, where it is clear that the oscillation frequency decreases for higher values of $k$. 

In Figures \ref{l=1ringdown} and \ref{l=2ringdown}, we see more oscillations than seen in Figure \ref{l=0ringdown}.  It is now easy to see that the damping rate, which is the magnitude of the slope of the maxima in the log graph, decreases as $k$ increases.  This is consistent with the data in Tables I and III.

To further check for the consistency between the numerically generated ringdown waveforms and QNM data provided in Tables I and III, we use the Prony method\cite{Prony} to extract the first ($n=0$) QNM frequency from the waveforms generated for the cases of $\frac{k}{2GM}=0.5$ for $l=0,1,2$ shown in Figures \ref{l=0ringdown}, \ref{l=1ringdown} and \ref{l=2ringdown} respectively.  
The results are $0.202199-0.193932i$, $0.542547-0.181305i$, and $0.896678-0.179637i$.  These are all consistent with the data shown in Tables I and III.

\section{Summary and Conclusion}
\label{Sec:conclusions}

We calculated the QNMs of a massless scalar field in the background of the regular black hole spacetime (\ref{PK-LE}) using the $6^{th}$ order WKB method and the AIM.  Both methods gave consistent results.  We also examined the ringdown waveform of this black hole and compared all our results with the Schwarzschild case.

The QNM frequencies of this regular black hole follow closely the QNM spectrum of a Schwarzschild black hole.  We found no modes with positive damping, which suggests such a regular black hole is stable against small massless scalar perturbations.
For the most part, we consider this black hole model a good Schwarzschild black hole mimicker.  The departure from general relativity is only observable when $\frac{k}{2GM}$ is not too small.  For macroscopic black holes, this requires the length scale $k$ to be very large compared to the Planck length.\footnote{We note that the authors of \cite{Babb} use the analytic technique developed by Motl and Neitzke in \cite{Motl} to show that, for small values of $\frac{k}{2GM}$, the high overtone QNMs of the regular black hole (\ref{PK-LE}) are almost the same as the Schwarzschild QNMs.  The spectrum begins to depart from the Schwarzschild spectrum when the damping $|\omega_I| \approx \frac{(2GM)^2}{k^3}$.  Our numerical results in this paper confirm that low overtone QNMs are also negligibly affected when $\frac{k}{2GM} << 1$.}  
We also showed that an increase in the magnitude of $k$ decreases the height of the QNM potential and gives oscillations with lower frequency and less damping.

 Most macroscopic regular black holes, including the one studied here, are well approximated by the Schwarzschild metric near the horizon as well as exterior to it. It is therefore not surprising that the corrections to the QNMs are small. On the other hand, the recent LQG black hole  proposed by Ashtekar, Olmedo and Singh\cite{AOS} has been shown\cite{Faraoni} to have some interesting but potentially unphysical properties near the horizon and asymptotically. It would be interesting to see how these features affect the QNM spectrum and the ringdown waveform. This work is in progress.

\vskip .5cm


\leftline{\bf Acknowledgments}
RGD thanks the William I. Fine Theoretical Physics Institute at the University of Minnesota where part of this research was conducted.  GK's work was funded in part by the Natural Sciences and Engineering Research Council of Canada. We also thank Wei-Liang Qian for sharing with us the code for the Prony method.



\def\jnl#1#2#3#4{{#1}{\bf #2} #3 (#4)}

\def\Zphys{{Z.\ Phys.} }
\def\jssc{{J.\ Solid State Chem.\ }}
\def\jpsJ{{J.\ Phys.\ Soc.\ Japan }}
\def\ptps{{Prog.\ Theoret.\ Phys.\ Suppl.\ }}
\def\PTP{{Prog.\ Theoret.\ Phys.\  }}
\def\LNC{{Lett.\ Nuovo.\ Cim.\  }}

\def\JMP{{J. Math.\ Phys.} }
\def\NPB{{Nucl.\ Phys.} B}
\def\NP{{Nucl.\ Phys.} }
\def\PLB{{Phys.\ Lett.} B}
\def\PL{{Phys.\ Lett.} }
\def\PRL{Phys.\ Rev.\ Lett.\ }
\def\PRA{{Phys.\ Rev.} A}
\def\PRB{{Phys.\ Rev.} B}
\def\PRD{{Phys.\ Rev.} D}
\def\PR{{Phys.\ Rev.} }
\def\PRe{{Phys.\ Rep.} }
\def\AP{{Ann.\ Phys.\ (N.Y.)} }
\def\RMP{{Rev.\ Mod.\ Phys.} }
\def\ZPC{{Z.\ Phys.} C}
\def\SCI{Science}
\def\CMP{Comm.\ Math.\ Phys. }
\def\MPLA{{Mod.\ Phys.\ Lett.} A}
\def\IJMPA{{Int.\ J.\ Mod.\ Phys.} A}
\def\IJMPB{{Int.\ J.\ Mod.\ Phys.} B}
\def\cmp{{Com.\ Math.\ Phys.}}
\def\JPA{{J.\  Phys.} A}
\def\CQG{Class.\ Quant.\ Grav.~}
\def\ATMP{Adv.\ Theoret.\ Math.\ Phys.~}
\def\AJP{Am.\ J.\ Phys.~}
\def\PRSA{{Proc.\ Roy.\ Soc.\ Lond.} A }
\def\ibid{{ibid.} }
\vskip 1cm

\leftline{\bf References}

\renewenvironment{thebibliography}[1]
        {\begin{list}{[$\,$\arabic{enumi}$\,$]}  
        {\usecounter{enumi}\setlength{\parsep}{0pt}
         \setlength{\itemsep}{0pt}  \renewcommand{\baselinestretch}{1.2}
         \settowidth
        {\labelwidth}{#1 ~ ~}\sloppy}}{\end{list}}


\end{document}